\documentclass[]{spie}  %>>> use for US letter paper
\usepackage{amsmath,amsfonts,amssymb}
\usepackage{graphicx}
\usepackage{booktabs}
\usepackage[colorlinks=true, allcolors=blue]{hyperref}

\title{Standardizing Medical Images at Scale for AI}

\author[a]{Callen MacPhee}

\author[a]{Yiming Zhou}

\author[b]{Koichiro Kishima}

\author[a,c]{Bahram Jalali}

\affil[a]{ECE Department, UCLA, 405 Hilgard Ave., Los Angeles, California 90095, USA}
\affil[b]{Pinpoint Photonics, Inc., 803 Marine Bldg., 4-23 Kaigan-dori, Naka, Yokohama 231-0002, Japan}
\affil[c]{Adventure Photonics, 889 N. Douglas Ave, El Segundo, California 90245, USA}

\begin{document} 
\maketitle

\begin{abstract}
Deep learning has achieved remarkable success in medical image analysis, yet its performance remains highly sensitive to the heterogeneity of clinical data. Differences in imaging hardware, staining protocols, and acquisition conditions produce substantial domain shifts that degrade model generalization across institutions. Here we present a physics-based data preprocessing framework based on the PhyCV (Physics-Inspired Computer Vision) family of algorithms, which standardizes medical images through deterministic transformations derived from optical physics. The framework models images as spatially varying optical fields that undergo a virtual diffractive propagation followed by coherent phase detection. This process suppresses non-semantic variability such as color and illumination differences while preserving diagnostically relevant texture and structural features. When applied to histopathological images from the Camelyon17-WILDS benchmark, PhyCV preprocessing improves out-of-distribution breast-cancer classification accuracy from $70.8\%$ (Empirical Risk Minimization baseline) to $90.9\%$, matching or exceeding data-augmentation and domain-generalization approaches at negligible computational cost. Because the transform is physically interpretable, parameterizable, and differentiable, it can be deployed as a fixed preprocessing stage or integrated into end-to-end learning. These results establish PhyCV as a generalizable \textit{data refinery} for medical imaging—one that harmonizes heterogeneous datasets through first-principles physics, improving robustness, interpretability, and reproducibility in clinical AI systems.
\end{abstract}

% Include a list of keywords after the abstract 
\keywords{Domain Shift, Pathology, Optical Physics, Physics-Inspired Algorithms, Computer Vision}

\section{INTRODUCTION}
\label{sec:intro}  % \label{} allows reference to this section

Artificial intelligence (AI) has become an indispensable tool for medical image analysis, driving advances in diagnostics, segmentation, and biomarker discovery~\cite{litjens2017survey, zhou2023deep}. Yet the performance and reliability of these models remain closely tied to the quality and consistency of their training data. Medical imaging data collected across institutions often differ in acquisition hardware, staining protocols, and storage conditions. These non-semantic variations introduce “batch effects”~\cite{schmitt2021hidden} that can limit generalization across clinical sites.

This challenge is reminiscent of the early oil industry, when crude oil from different sources varied widely in composition and usability. The development of refining standards and processing routines transformed these disparate inputs into a consistent and reliable resource. In much the same way, although medical data is abundant and information rich, it requires systematic refinement to ensure interoperability and consistency. Each institution contributes a slightly different form of data, and without standardization routines to harmonize them, even well-trained models may produce inconsistent results.

Creating such a ``data refinery” does not necessarily demand large-scale infrastructure but rather practical, efficient routines that can align and normalize data with minimal complexity. These processes must be fast, transparent, and designed to preserve meaningful biological variation while reducing irrelevant differences. Establishing this type of lightweight standardization could help make medical AI more robust, reproducible, and generalizable across diverse healthcare settings.

\subsection{Prior Art}
To address data heterogeneity in the medical domain, various approaches have been proposed, which can be broadly categorized into two groups: data augmentation and data standardization. Data augmentation generates additional training samples that randomize only spurious variations while keeping the robustly predictive factors~\cite{gao2023out,nguyen2023contrimix}. This approach enlarges existing training data and increases the generalization ability of AI models in unseen domains. However, data augmentation incurs additional computational costs and may require larger models. Data standardization, on the other hand, serves as a preprocessing step before training AI models, aiming to eliminate data heterogeneity by aligning the data distribution of training set and test set. For example in histopathological images, stain color normalization is often used to align the distribution of stain colors~\cite{ruifrok2001quantification, macenko2009method, vahadane2016structure}. However, existing normalization methods require extracting specific features or training dedicated models, which can be challenging when processing large volumes of data at scale.  

\begin{figure}[t]
    \centering
    \includegraphics[width=\linewidth]{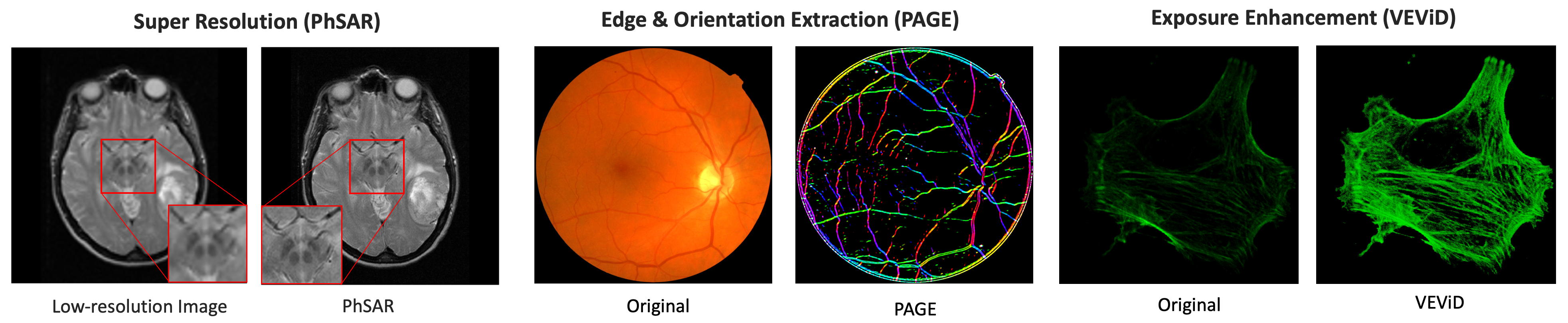}
    \caption{\textbf{Examples of physics-inspired computer vision algorithms in PhyCV.} 
    The PhyCV framework includes modules for diverse imaging tasks: super-resolution (PhSAR) enhances structural detail in low-resolution MRI scans; 
    edge and orientation extraction (PAGE) reveals vascular and textural organization in retinal imagery; 
    and exposure enhancement (VEViD) restores visibility in low-contrast microscopy images. 
    Together, these algorithms demonstrate the versatility of physics-based models for image refinement and feature extraction across modalities.}
    \label{fig:phycv_examples}
\end{figure}

\subsection{Data Standardization at Scale with PhyCV}

The Physics-Inspired Computer Vision (PhyCV) family of algorithms are a collection of computationally efficient approaches to texture and edge detection, low-light image enhancement, and dynamic signal analysis. Notably, the framework has demonstrated superior performance in extracting relevant features across various medical imaging applications, such as single molecule biological imaging~\cite{ilovitsh2016phase}, retina vessel extraction~\cite{challoob2020local}, and super-resolution of MRI images~\cite{he2019fast}, as shown in Figure \ref{fig:phycv_examples}. These applications highlight PhyCV's potential as a promising tool for data refinement in the medical domain. In this paper, we focus on how PhyCV can be utilized to standardize pathology images at scale to enable more accurate and robust cancer detection. We propose a simple yet effective preprocessing pipeline that leverages PhyCV to align the data distribution of pathology images from different sources, thereby reducing the impact of data heterogeneity on AI model performance. Our method is computationally efficient and can be easily integrated into existing deep learning pipelines for medical image analysis.

The proposed framework employs PhyCV as a physics-based data refinement and standardization stage that unifies heterogeneous inputs into a consistent feature representation. During training, data originating from multiple clinical sources, each with different illumination, staining, and imaging characteristics, are refined through PhyCV to form a standardized feature representation on which a deep neural network is trained. The same refinement protocol is applied prior to inference, ensuring that new data are projected into the same unified representation established during training. Because PhyCV is grounded in phase detection and spectral phase processing, it inherently performs self-equalization of contrast and illumination variations. This natural equalization, emerging directly from the physics of the spectral phase modulation and coherent detection, provides a strong motivation for adopting PhyCV as a physically interpretable and domain-robust preprocessing framework.

We validate this approach using histopathology images from the Camelyon17-WILDS dataset, a benchmark designed to test out-of-distribution generalization. PhyCV preprocessing yields significant improvements in classification accuracy across institutions while introducing minimal computational overhead. By standardizing medical images at the feature level, this framework offers a scalable pathway toward more robust, transferable, and interpretable AI systems in clinical imaging.
\section{PhyCV}

\subsection*{Roots of PhyCV in Photonic Time Stretch}
The physical foundations of PhyCV trace back to the principles of photonic time stretch, a technique originally developed to capture ultrafast events beyond the bandwidth limits of electronic digitizers \cite{jalali1998photonic}. In photonic time stretch, an optical pulse carrying information about a high-speed phenomenon upon its spectrum is propagated through a dispersive medium, such as an optical fiber, where group-velocity dispersion imposes a frequency-dependent delay \cite{zhou2022unified}. Dispersion stretches the temporal signal based on the spectral profile, thereby slowing down the signal for real-time digitization by conventional analog-to-digital converters. The elegance of this process lies in its use of inherent frequency-dependent phase modulation in chromatic dispersion to redistribute information in time.

PhyCV generalizes this same principle to computational imaging: instead of stretching time, it performs a virtual optical propagation that applies a spectral phase kernel in the spatial-frequency domain \cite{jalali2022vevid, asghari2015pst}. This operation emulates diffraction through a virtual medium, redistributing spatial information analogously to how dispersion redistributes temporal frequencies. When the propagated field is coherently detected, the resulting phase map emphasizes structural and textural features while inherently equalizing illumination and color variations. Thus, PhyCV can be viewed as a computational analog to photonic time stretch which transfers the physics of dispersive propagation from the temporal domain of optical instrumentation to the spatial diffractive domain for computer vision. Notably, the use of spectral phase modulation and coherent detection leads to natural intensity, color, and saturation equalization described mathematically below.

\begin{figure}[t]
    \centering
    \includegraphics[width=\linewidth]{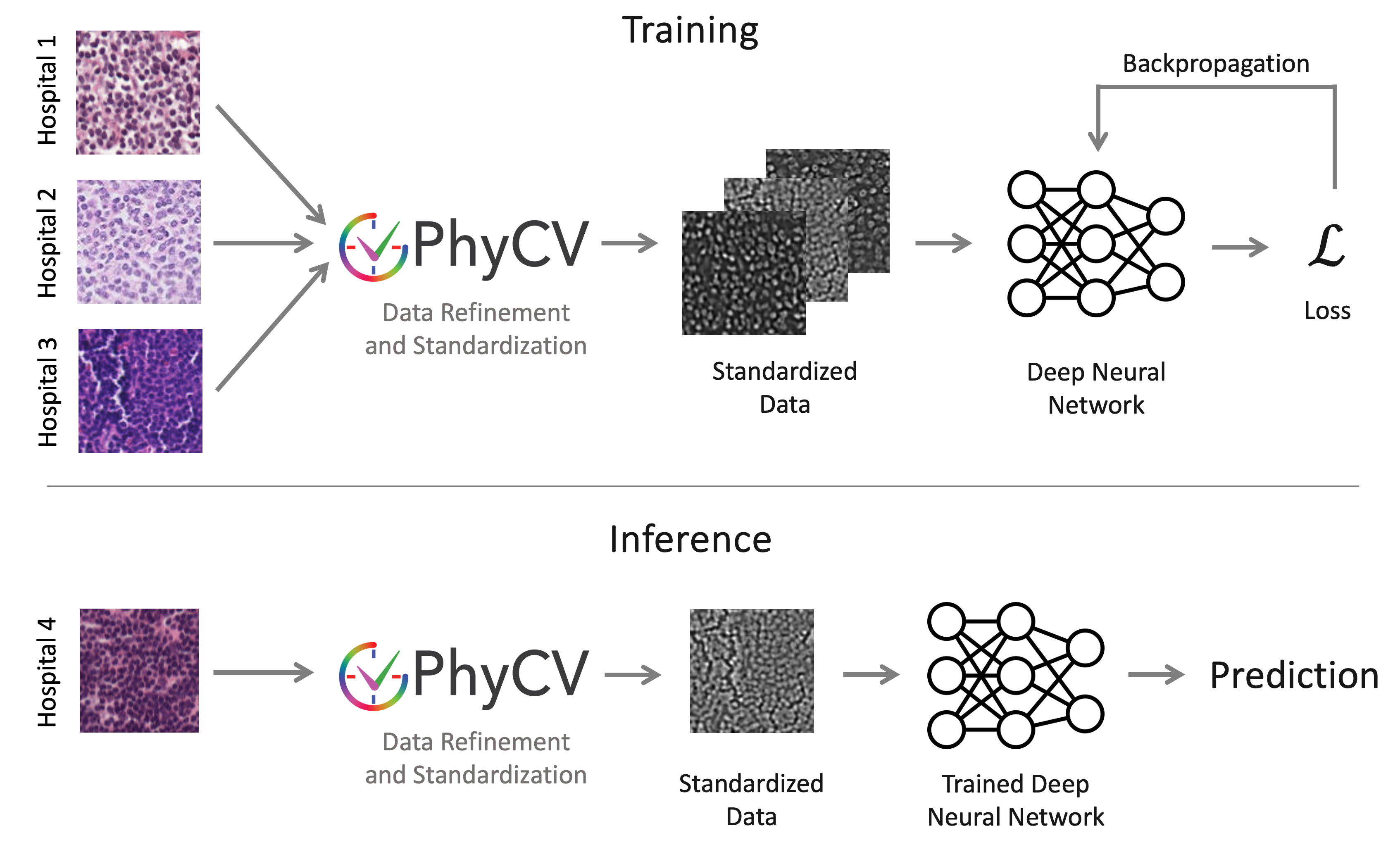}
    \caption{\textbf{Conceptual overview of PhyCV for data refinement and standardization.}
    During training (top), heterogeneous data from multiple hospitals are refined into standardized feature representations before being used to train a neural network. 
    During inference (bottom), unseen data undergo the same refinement to ensure consistent predictions.}
    \label{fig:concept}
\end{figure}

\subsection*{Mathematical Foundations of PhyCV}

The algorithms within the PhyCV family emulate the propagation of light through a two-dimensional diffractive medium with engineered optical properties, followed by coherent (phase) detection. Within this medium, propagation is parameterized via a spectral phase kernel applied in the frequency domain, transforming a real-valued image into a complex-valued optical field. Subsequent coherent detection in the spatial domain extracts phase information that encodes salient structural and textural features of the input image.

Let the input image be represented by a real-valued function \( E_i(x, y) \).  
The corresponding two-dimensional spatial spectrum is obtained via the Fourier transform

\begin{equation}
\tilde{E_i}(k_x, k_y) = \mathcal{F}\{ E_i(x, y) \},
\label{eq:fourier}
\end{equation}
\newline
where \( k_x \) and \( k_y \) denote spatial frequency coordinates.  
Virtual propagation through a  dielectric of length \( z \) imparts a frequency-dependent phase shift described by a spectral phase kernel \( \phi(k_x, k_y) \).  
The modulated spectrum becomes

\begin{equation}
\tilde{E_o}(k_x, k_y) = \tilde{E_i}(k_x, k_y) \, e^{\, -i\phi(k_x, k_y)}.
\label{eq:spec_phase_kernel}
\end{equation}
\newline
The inverse Fourier transform then reconstructs the propagated field in the spatial domain:

\begin{equation}
E_o(x, y) = \mathcal{F}^{-1}\{ \tilde{E_o}(k_x, k_y) \}
\label{eq:ifft_field}
\end{equation}
\newline
The resulting complex field \( E_o(x, y) \) can be interpreted as the optical response of the input image after experiencing virtual dispersion. The output of the algorithm is the spatial phase of this signal, represented by an array of phase pixels or ``phixels''
\newline
\begin{equation}
\psi(x, y) = \tan^{-1}\!\left(
\frac{Im\{E(x, y)\}}{Re\{E(x, y)\}}
\right),
\label{eq:phase_detect}
\end{equation}
\newline
and serves as the output of PhyCV algorithms.  
A normalization step \( \mathcal{N}(\cdot) \) can be applied to scale the resulting phase or magnitude map to a fixed numerical range suitable for digital image representation:

\begin{equation}
I_{\mathrm{out}}(x, y) = \mathcal{N}\big( \psi(x, y) \big).
\label{eq:normalization}
\end{equation}

\subsection*{Small-Phase Approximation and Feature Equalization}

In PhyCV implementations, the spectral phase typically satisfies \( |\phi(k_x, k_y)| \ll 1 \), allowing approximation via a first-order Taylor expansion:

\begin{equation}
e^{\, -i\phi(k_x, k_y)} \approx 1 + i\phi(k_x, k_y).
\label{eq:small_phase}
\end{equation}
\newline
Substituting Eq.~(\ref{eq:small_phase}) into Eq.~(\ref{eq:ifft_field}) gives

\begin{align}
E_o(x, y) 
&\approx \mathcal{F}^{-1}\{ \tilde{E_i}(k_x, k_y) \} 
- i \, \mathcal{F}^{-1}\{ \tilde{E_i}(k_x, k_y)\phi(k_x, k_y) \} \nonumber \\
&= E_i(x, y) - i \, \mathcal{G}\{ E_i(x, y) \},
\label{eq:spatial_approx}
\end{align}
\newline
where \( \mathcal{G}\{\cdot\} \) is a spatial-domain  operator related to \( \phi(k_x, k_y) \).  
For the quadratic kernel in the case of group-velocity dispersion, \( \mathcal{G}\{ E_i(x, y) \} \) corresponds to a Laplacian-like operator, i.e.,

\begin{equation}
\mathcal{G}\{ E_i(x, y) \} \propto \nabla^2 E_i(x, y).
\label{eq:laplacian}
\end{equation}
\newline
The detected phase can therefore be approximated by

\begin{equation}
\psi(x, y) = 
\tan^{-1}\bigg( \frac{Im\{E_o(x, y)\}}{Re\{E_o(x, y)\}}\bigg) 
\propto
\frac{\mathcal{G}\{E_i(x, y)\}}{E_i(x, y)} 
\propto 
\frac{\nabla^2 E_i(x, y)}{E_i(x, y)}.
\label{eq:phase_small}
\end{equation}
\newline
Equation~(\ref{eq:phase_small}) reveals that the PhyCV operation exhibits behavior similar to an equalized Laplacian, enhancing edges and fine textures while minimizing the effect of global intensity or staining variations.

\subsection*{Integration with Machine Learning}

The PhyCV preprocessing described by Eqs.~(\ref{eq:fourier})–(\ref{eq:phase_small}) is deterministic and fully differentiable.  
In Figure \ref{fig:concept}, we see the use of the PhyCV framework within the machine learning pipeline. In the training phase (top), data from multiple hospitals, each exhibiting different illumination, staining, and imaging conditions, are passed through PhyCV for data refinement and standardization. The resulting standardized images are then used to train a deep neural network, with gradients and loss computed through standard backpropagation. During inference (bottom), unseen data from a new hospital undergo the same PhyCV-based refinement before being processed by the trained network for prediction. This ensures that the inference data are represented in the same standardized domain as the training data. Because PhyCV operates on principles of phase detection and spectral phase processing, it inherently performs self-equalization of intensity and contrast variations, providing a physics-grounded motivation for its use as a robust data standardization framework.

\section{Results}

We first examine PhyCV’s performance in a controlled setting designed to emulate a common data refinery challenge: non-uniform illumination \cite{aslan2025impact}. This study isolates how the algorithm responds to structured variations in image intensity, allowing us to assess its ability to remove non-semantic artifacts while preserving essential structural information. Following this controlled analysis, we evaluate PhyCV on a large-scale pathology dataset collected across multiple clinical sites, where the diversity in staining, acquisition hardware, and imaging protocols presents a more complex and realistic test of data standardization.

\subsection*{Evaluation on Simulated Non-Uniform Illumination}

\begin{figure}[htbp]
    \centering
    \includegraphics[width=\linewidth]{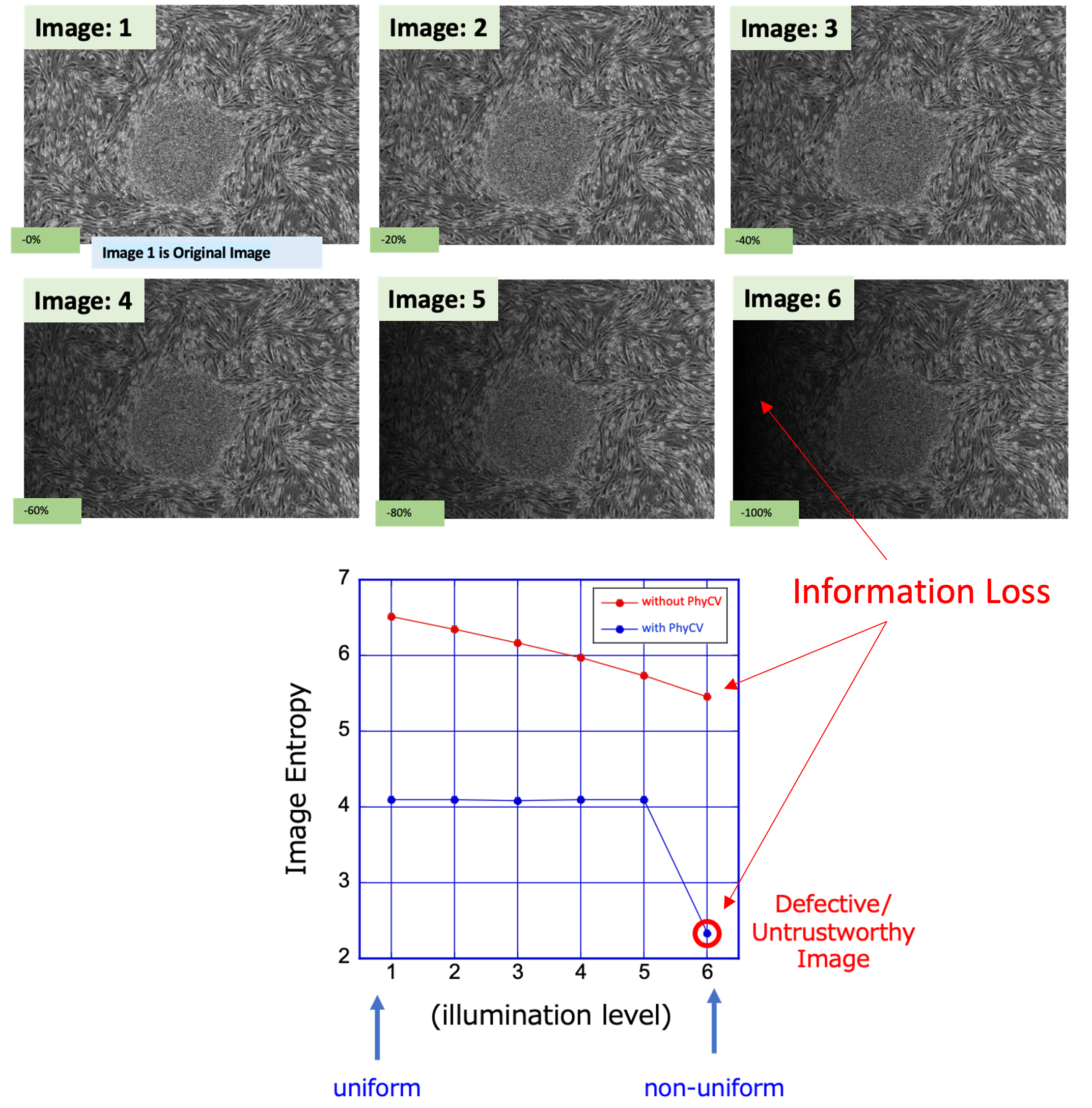}
    \caption{\textbf{PhyCV improves robustness under non-uniform illumination.} 
    Top: Non-uniform illumination is studied by considering 6 levels of illumination (linear) on a single induced pluripotent stem cell (iPS) test image picture from  \cite{okamoto2011induction}. For each of the 6 illumination levels, center region is used for texture analysis. Bottom: corresponding image entropy showing that PhyCV maintains information content even under severe illumination degradation. When information is fully lost in the case of image 6, there is a noteacble loss in the PhyCV output, demonstrating that it can act as a delineator for defective or untrustworthy images.}
    \label{fig:illumination}
\end{figure}

Non-uniform illumination presents a major challenge in automated image analysis, as it introduces spatial bias that can distort texture and intensity-based measurements. Variations in lighting conditions, whether from optical misalignment, uneven illumination sources, or sample thickness, can introduce spatial intensity gradients that confound automated analysis pipelines. These differences do not alter the underlying biological content but can produce systematic biases in feature extraction, segmentation, and texture quantification. To evaluate PhyCV’s ability to mitigate such effects, we designed a controlled study in which illumination levels were systematically varied to simulate realistic degradations. This experiment provides a quantitative framework to assess how well PhyCV can “refine” data by removing non-semantic illumination artifacts while preserving essential structural information.

Figure \ref{fig:illumination} demonstrates how PhyCV mitigates this effect through its inherent illumination-invariant transformation. Six levels of illumination were applied to a pluripotent stem cell image to simulate increasingly non-uniform lighting conditions. Conventional analysis showed a progressive loss of information content, as reflected in the decreasing image entropy (red curve). In contrast, PhyCV maintained stable entropy across illumination levels (blue curve), indicating preservation of essential image structure and texture despite varying brightness conditions. When illumination became fully corrupted (image 6), the PhyCV output exhibited a distinct drop in information content, effectively identifying the image as defective or unreliable. This behavior suggests that PhyCV not only standardizes non-uniform illumination but can also serve as an intrinsic quality control mechanism—preserving information where possible and flagging instances where data integrity has been compromised.

\begin{figure}[htbp]
    \centering
    \includegraphics[width=\linewidth]{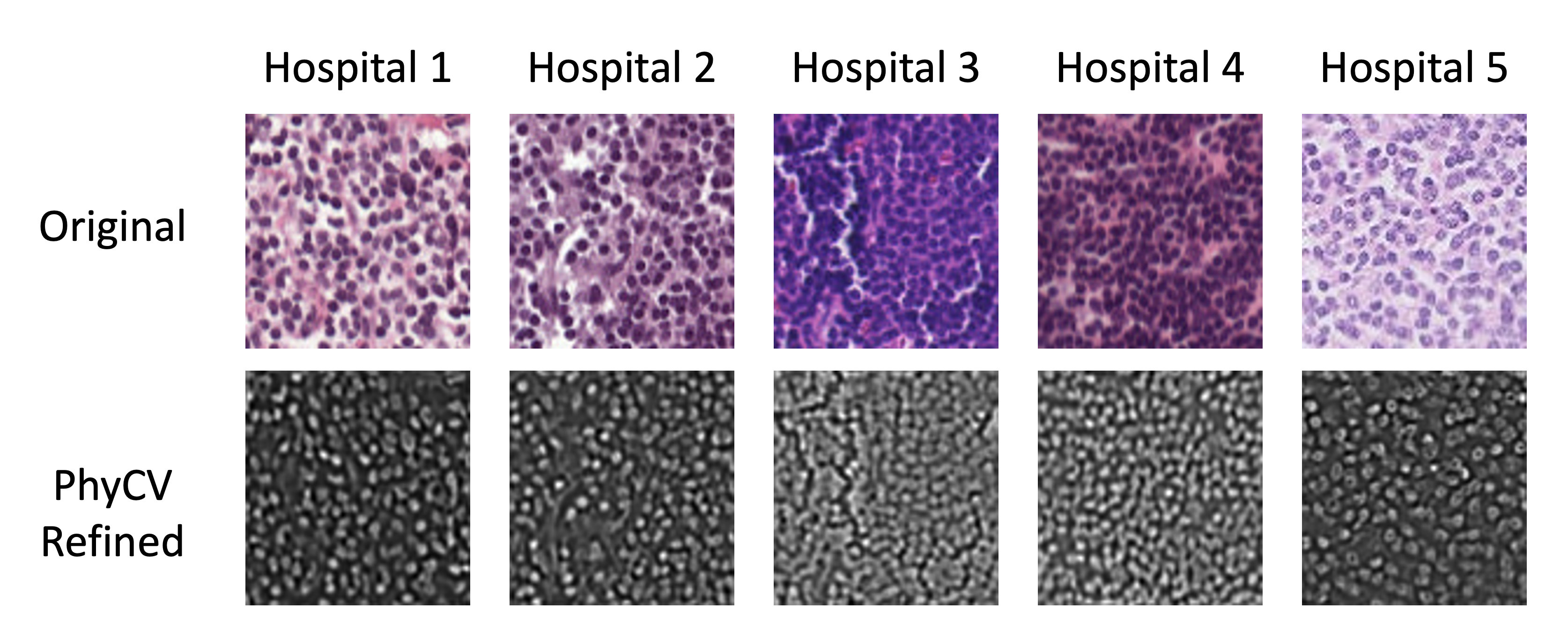}
    \caption{\textbf{Cross-institutional variability and PhyCV-based refinement.} 
    Example histopathology patches from five hospitals showing differences in staining and contrast (top). 
    After PhyCV refinement (bottom), features are standardized, enhancing consistency across sites while preserving tissue structure.}
    \label{fig:refinement}
\end{figure}

\subsection*{Evaluation on Histopathology Data}

Next, we demonstrate effectiveness of PhyCV standardizing histopathological images at scale for more accurate and robust breast cancer detection. We report the results on the Camelyon17-WILDS dataset consists of 50 whole-slide images (WSIs) of breast cancer metastases in lymph node sections sourced from 5 different hospitals in the Netherlands~\cite{bandi2018detection, koh2021wilds}. Specifically, 10 WSIs are sourced from each of the 5 hospitals. A total of 450,000 patches, each of size $96\times96$, are extracted from these 50 WSIs. The label of each patch serves as a binary indicator denoting the presence of tumor tissue within the central $32\times32$ region. The training set consists of 302,436 patches extracted from 30 WSIs, with 10 WSIs originating from each of the three hospitals designated for training. The in-distribution (ID) validation set contains 33,560 patches derived from the same 30 WSIs utilized in the training set, while the out-of-distribution (OOD) validation set comprises 34,904 patches acquired from 10 WSIs belonging to the fourth hospital. Notably, these WSIs are distinct from those in the other splits. The out-of-distribution (OOD) test set comprises 85,054 patches obtained from 10 WSIs affiliated with the fifth hospital, chosen due to the unique visual distinctiveness of its patches.
  
For PhyCV, we use official Phase Stretch Transform (PST) implementation from the open-sourced PhyCV library~\cite{zhou2023phycv} with the following parameters: $S=0.3$, $W=15$, $\sigma_{LPF}=0.15$. Note that the thresholding and morphological operations are not applied here, so the output of PST is an “analog” feature map instead of a “digital” binary edge map. This is done intentionally as it keeps richer semantic information for the neural network training.
  
Regarding the AI model, a DenseNet-121 model~\cite{huang2017densely} is trained from scratch on the Camelyon17-WILDS dataset from scratch with a learning rate of $10^{-3}$, $L_2$ regularization strength of $10^{-2}$, a batch size of 32 and SGD with momentum (set to 0.9). Training comprises 5 epochs, and results for each preprocessing technique are aggregated over 10 random seeds. The implementation follows the official open-sourced codebase~\cite{koh2021wilds}.

\subsection*{Quantitative Performance}

Table \ref{tab:camelyon_results} presents the classification accuracy on the out-of-distribution (OOD) validation and test sets of the Camelyon17-WILDS dataset for various methods. The results are aggregated over 10 random seeds, with standard deviations shown in parentheses. It is important to note that all methods reported here use the same data split and neural network architecture (DenseNet121) for a fair comparison.

The baseline method, Empirical Risk Minimization (ERM), uses SGD with momentum as the optimizer and doesn't apply any data augmentation or standardization techniques. PAIR~\cite{chen2022pareto} and Fish~\cite{shi2021gradient} are novel training algorithms designed to improve domain generalization performance. LISA~\cite{yao2022improving}, ERM with target augmentation~\cite{gao2023out}, and ContriMix~\cite{nguyen2023contrimix} are data augmentation techniques to enhance the model performance in unseen domains. Our approach distinguishes itself among the existing methods as a data standardization technique. Unlike other methods that introduce extra augmented data or modify training algorithms, PhyCV is a straightforward preprocessing step that significantly improves out-of-distribution classification performance. By standardizing the input data, PhyCV helps the model learn more robust and generalizable features, leading to better performance on unseen domains.

\begin{table}[t]
\centering
\caption{\textbf{Out-of-distribution classification accuracy (\%) on the Camelyon17-WILDS dataset.}  
Mean $\pm$ standard deviation computed across $10$ random seeds.  All methods use the same network (DenseNet-121) and data split for fair comparison.}
\label{tab:camelyon_results}
\begin{tabular}{lcc}
\toprule
\textbf{Method} & \textbf{OOD Validation} & \textbf{OOD Test} \\
\midrule
ERM (baseline)           & $85.8 \, (1.9)$ & $70.8 \, (7.2)$ \\
PAIR                     & $84.3 \, (1.6)$ & $74.0 \, (7.2)$ \\
Fish                     & $83.9 \, (1.2)$ & $74.7 \, (7.1)$ \\
LISA                     & $81.8 \, (1.2)$ & $77.1 \, (6.9)$ \\
ERM + target augmentation & $92.7 \, (0.7)$ & $92.1 \, (3.1)$ \\
ContriMix                & $91.9 \, (0.7)$ & $94.6 \, (1.2)$ \\
\textbf{PST (ours)}      & $\mathbf{90.3 \, (0.8)}$ & $\mathbf{90.9 \, (2.4)}$ \\
\bottomrule
\end{tabular}
\end{table}

\subsection*{Qualitative Observations}

Visual inspection of the PhyCV-processed images in Figure \ref{fig:refinement} confirms that the transform standardizes illumination while preserving diagnostically relevant structures. 
Compared to raw data, the PhyCV outputs exhibit enhanced edge definition and reduced inter-hospital variability.  
The resulting feature maps provide neural networks with a consistent statistical input distribution, which explains the observed improvement in OOD accuracy.  
Importantly, this standardization is achieved without introducing synthetic artifacts or requiring reference slides, differentiating PhyCV from conventional stain-normalization approaches. Because PhyCV operates via a single forward and inverse Fourier transform with a low-order phase kernel, its computational cost is negligible compared with deep-network training or inference.   

\subsection*{Summary of Findings}

Across all experiments, PhyCV preprocessing demonstrably improves the robustness and cross-domain generalization of deep learning models for pathology image analysis.  
By removing non-semantic variability at the data level rather than at the model level, the PhyCV framework enables consistent performance across heterogeneous clinical datasets while maintaining minimal computational overhead.  
These results validate the concept of a physics-based data refinery as a practical and scalable strategy for standardizing medical imagery at scale.

\section{Discussion}

Deep learning models in medical imaging have traditionally depended on large, carefully curated datasets to achieve robust performance across clinical environments.  
However, such dependence is fundamentally at odds with the heterogeneity of real-world medical data, where acquisition conditions, device characteristics, and tissue preparation protocols vary widely across institutions.  
The results presented here demonstrate that PhyCV preprocessing provides a physically interpretable and computationally efficient means of mitigating this variability prior to learning.  
By encoding image content through a virtual optical propagation and coherent phase detection process, PhyCV effectively normalizes illumination, staining, and texture differences while preserving diagnostically relevant features such as glandular structure and cellular morphology.

\subsection*{The Role of Physics-Based Data Refinement}

The PhyCV framework introduces a conceptual shift from data augmentation and model regularization toward physics-based data refinement.  
Rather than training models to learn invariance from vast, heterogeneous data, PhyCV explicitly imposes invariance through deterministic, physically motivated transformations.  
This paradigm parallels industrial refinement pipelines by transforming raw, inconsistent input materials into standardized, high-quality products.  
Applied to medical imaging, this approach ensures that neural networks receive data with reduced domain-specific noise and enhanced interpretability, effectively decoupling diagnostic content from acquisition variability.

Unlike empirical normalization methods, PhyCV preprocessing derives directly from first principles of electromagnetic propagation.  
This physical foundation ensures that the transformation is globally consistent, parameterizable, and independent of dataset-specific references.  
Because PhyCV operates deterministically, it preserves reproducibility across institutions and implementations which is an essential property for regulatory compliance and clinical deployment.  
Furthermore, the differentiable nature of its transform allows it to be integrated into hybrid “physics-learning’’ architectures, in which spectral phase kernels are optimized jointly with network parameters to adapt to particular imaging modalities.

\subsection*{Implications for Clinical Translation}

The improvements in out-of-distribution accuracy observed on the Camelyon17-WILDS benchmark indicate that physics-based preprocessing can significantly narrow the generalization gap between institutions without increasing model complexity or data requirements.  
This scalability is critical for clinical translation, where annotated datasets remain limited and imaging systems differ across sites.  
The negligible computational overhead of PhyCV preprocessing further supports its integration into large-scale or real-time clinical pipelines.  
Because PhyCV outputs are phase-based and human-interpretable, they also provide a natural bridge between conventional image processing and explainable AI, enabling joint visualization of physical and learned features.

The data refinery concept also extends beyond histopathology.  
Virtual diffraction and phase-based standardization can, in principle, be applied to radiological, endoscopic, or microscopic modalities, where acquisition heterogeneity and contrast variability similarly limit AI deployment.  
In these contexts, PhyCV provides a unified front-end transformation that harmonizes inputs without compromising downstream model flexibility.

\subsection*{Limitations and Future Directions}

While PhyCV significantly improves robustness to inter-institutional variability, its performance depends on the selection of appropriate kernel parameters (\(S\), \(W\), and \(\sigma_{\mathrm{LPF}}\)) and phase scaling.  
Although these parameters are physically interpretable, systematic tuning may be required to optimize results across modalities.  
In addition, while the current implementation operates on two-dimensional image patches, extension to volumetric data and temporal sequences could further enhance its applicability to three-dimensional microscopy or dynamic imaging tasks.  
Future work will explore adaptive kernel optimization, where the spectral phase profile is learned end-to-end while remaining constrained by physical priors—merging physics-based interpretability with the flexibility of deep learning.

\subsection*{Outlook}

The findings presented here establish PhyCV as a generalizable and scalable framework for data standardization in medical imaging.  
By bridging the gap between optical physics and machine learning, PhyCV redefines data preprocessing as a form of computational optics–one that standardizes information before it reaches the learning stage.  
This shift toward physics-informed data refinement has implications not only for histopathology but for any domain where data heterogeneity constrains model reliability.  
As the volume of clinical imaging data continues to grow, such physics-grounded preprocessing may become a foundational step in building robust, transferable, and interpretable medical AI systems.

\bibliography{report} % bibliography data in report.bib
\bibliographystyle{spiebib} % makes bibtex use spiebib.bst

\end{document}